\documentclass[aps,prl,twocolumn,showpacs,letterpaper,superscriptaddress]{revtex4-1}

\usepackage[colorlinks=true, citecolor=blue, linkcolor=blue, urlcolor=blue]{hyperref}

\usepackage{graphicx,dcolumn,longtable,epsfig}
\usepackage[usenames]{color}
\usepackage{amssymb}
\usepackage{amsmath}
\usepackage{epstopdf}
\usepackage{bm}
\usepackage{footnote}
\usepackage{float}
\usepackage{subfigure}
\usepackage{color}

\input epsf.tex

\def\atom#1#2{$^{#1}$#2}
\def\bea{\begin{eqnarray}}
\def\eea{\end{eqnarray}}
\def\be{\begin{equation}}
\def\ee{\end{equation}}

\begin{document}

\title{The Role of Interspecies Interactions in the Preparation of a Low-entropy Gas of Polar Molecules in a Lattice}
\author{A. Safavi-Naini }
\author{M.~L.~Wall}
\author{A. M. Rey}
\affiliation{JILA, NIST and Department of Physics, University of Colorado, 440 UCB, Boulder, Colorado 80309, USA }
\begin{abstract}
{The preparation of a quantum degenerate gas of heteronuclear molecules has been an outstanding challenge. We use path integral Quantum Monte Carlo simulations to understand the role of interactions and finite temperature effects in the  protocol currently employed to adiabatically prepare a low-entropy gas of polar molecules in a lattice starting from an ultracold Bose-Fermi mixture. We find that interspecies interactions affect the final temperature of the mixture after the adiabatic loading procedure and detrimentally limit the molecular peak filling. Our conclusions are in agreement with recent experimental measurements~\cite{Moses_Covey_15} and therefore are of immediate relevance for the myriad experiments that aim to form molecules from dual-species atomic gases. }
\end{abstract}

\maketitle

\begin{figure}[t]
\includegraphics[width=0.35\textwidth]{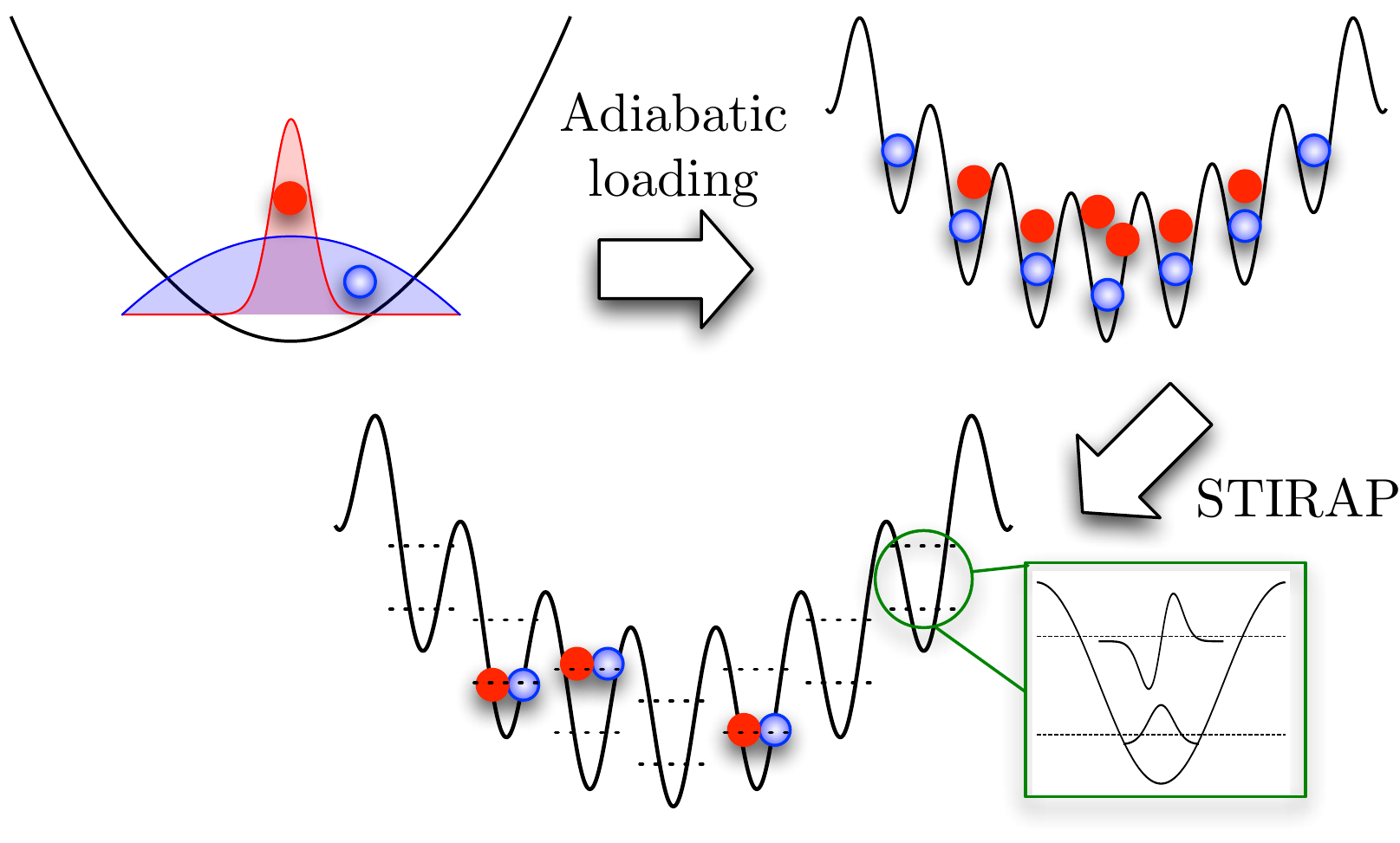}
\caption{(Color online) \emph{Production of ground state molecules from a dual-species mixture.}  (Top line) A finite-temperature dual-species gas of bosonic species A atoms (red) and fermionic species B atoms (blue), is adiabatically loaded from a harmonic trap into an optical lattice in the presence of interspecies interactions.  The initial temperature and interactions determine the temperature, density overlap, and peak filling of the mixture.  (Bottom line) Following STIRAP, only those sites with exactly one atom of each species are converted to a ground state molecule, with some probability to excite the molecule to a higher lattice band (schematic wavefunctions in green box).}
\label{fig:fig0}
\end{figure}

{\it Introduction.}
Polar molecules, interacting via long-range and anisotropic dipolar interactions, hold great promise as quantum simulators hosting exotic quantum phases~\cite{Carr_Demille_09,doi:10.1021/cr2003568}, as well as a diverse range of phenomena, ranging from quantum magnetism~\cite{Wallreview}, to many-body localization~\cite{MBL}, to synthetic spin-orbit coupling~\cite{Chiron, Buechler}. A necessary ingredient for simulating these these behaviors is to the ability to reach low entropy conditions. However, despite the rapid experimental progress, a reliable method to form such a state has remained out of reach~\cite{yan:observation_2013,PhysRevLett.112.070404,hazzard:many-body_2014}. The two main obstacles to realizing a low-entropy state are the inapplicability of standard atomic cooling techniques to polar molecules and the requirement to suppress chemical reactions. A proposed solution to these problems is to form the  molecules directly in a deep optical lattice through association after optimally loading a degenerate atomic gas mixture~\cite{Ni_Ospelkaus_08,Cho_McCarron_11,C1CP21769K,PhysRevA.85.032506,Takekoshi_14,
PhysRevLett.109.085301,PhysRevA.86.021602,PhysRevA.89.020702,1505.00473,Quemener_Julienne_12}. This protocol is shown schematically in Fig.~\ref{fig:fig0}. However, in order for this scheme to be a reliable pathway, it is necessary to have a high atom/molecule conversion efficiency by
creating a large region where the densities of the two atomic species overlap and correspond to exactly one atom of each species per lattice site. Temperature, interactions, and loading conditions can significantly  limit the achievement of  this requirement.

In this Letter, we use path integral Monte Carlo simulations (QMC) based on the worm algorithm~\cite{Worm}, as well as its two-worm extension, to investigate the effects of finite temperature and interspecies interactions during adiabatic loading from a dipole trap into an optical lattice.  We show that the final temperature of the lattice system following adiabatic loading depends strongly on the strength and sign of interspecies interactions, and can be far from the ideal, zero-temperature regime. In contrast to previous studies, which do not treat the adiabatic loading procedure and find that attractive interspecies interactions enhance the on-site densities \cite{Damski_Creation,Freericks_efficiency}, our analysis predicts that both attractive and repulsive interactions can lead to substantial depletion of on-site densities of the bosonic species and takes into account the contributions of dimensionality, effective mass imbalance, and quantum statistics~\cite{Jaksch_Venturi, Pollet_Kollath_1,Pollet_Kollath_2}. Additionally, we discuss the impact of imperfections in stimulated Raman Adiabatic Passage (STIRAP), which is used to convert Feshbach molecules to ground state molecules, on the ground state molecule production \cite{Ospelkaus_Peer_08,Ni_Ospelkaus_08}. We study the adiabaticity of this protocol, and provide approximate analytical formulas which can be used to determine the probability of promotion of molecules to higher bands during the STIRAP procedure. Since molecules in higher bands have a much larger tunneling rate, an appreciable higher band population can greatly impact dynamics involving molecular motion. We find that for experimentally relevant parameters there is no significant population transfer, provided the Lamb-Dicke criterion is satisfied. Our conclusions are corroborated by the recent experimental observations reported in~\cite{Moses_Covey_15} and thus of fundamental importance for ongoing experimental efforts to achieve a high-filling lattice gas of ground state polar molecules.

{\it Single-component gases in the lattice}
We begin by studying trapped single-species gases of bosons or fermions. We assume the bosons are  trapped in a  deep lattice where they can be described by the Bose-Hubbard Hamiltonian with an additional external harmonic confinement:
\begin{align}
\label{eq:Hsingle}
\nonumber H&=-\sum_{\langle i\,j\rangle}J_{\mathrm{A}}(a_{i,\rm A}^\dagger a_{j,\rm A}+\mathrm{h.c.})+\frac{U}{2}\sum_{i}n_{i,\rm A}(n_{i,\rm A}-1)\\
&-\sum_{i}\mu_{i,\rm A} n_{i,\rm A}.
\end{align}
Here, $a_{i,\rm A} ^\dagger$ ($a_{i, \rm A}$) is the bosonic creation (annihilation) operator for species ``A" and $n_{i,\rm A}=a_{i,\rm A}^\dagger a_{i,\rm A}$.  The first and second terms in the Hamiltonian Eq.~\eqref{eq:Hsingle} are the tunneling $J_{\mathrm{A}}$ and the on-site interaction, assumed repulsive with strength $U>0$, respectively. Finally, $\mu_{i,\rm A}=\mu-\sum_{\xi=x,y,z} w_\xi \xi_i^2$, where $w_\xi$ and $\xi_i$ are the strength of harmonic confinement and the coordinate of site $i$ along axis $\xi$, respectively, and $\mu$ is the chemical potential.

%, which, in the absence of harmonic trapping, sets the number of particles of type A in the system. Here $\beta=x,\,y,\,z$ are the trap axes, $w_\beta$ is the strength of harmonic confinement along the axes, and $(x_0, y_0, z_0)$ is the coordinate of the trap center.

In order to facilitate comparisons with the experimental data in~\cite{Moses_Covey_15}, we will use \atom{87}{Rb} for species A, but we stress that our conclusions apply for generic bosonic species described by the Bose-Hubbard model. The effective external harmonic confinement includes the additional confinement due to the curvature of the lattice beams. The A atoms are assumed  to experience a lattice potential $V_0=20 E_{A}$, where $E_{A}$ is the recoil energy of the A atoms.

\begin{figure}[t]
\includegraphics[trim={1cm 1.5cm 5cm 0cm}, clip,width=0.5\textwidth]{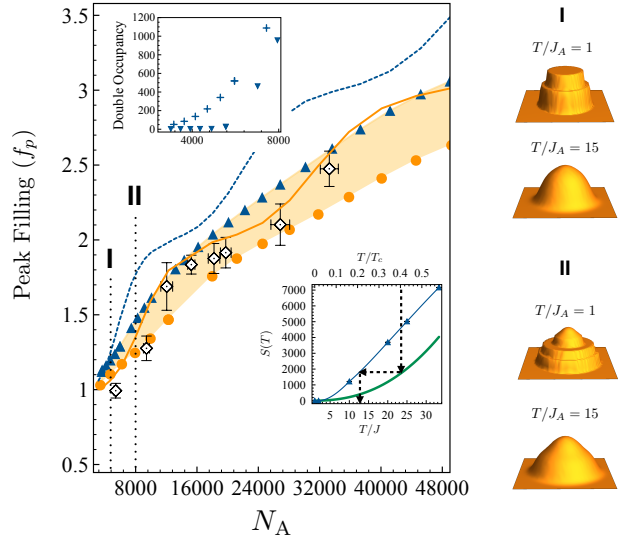}
%\caption{(Color online) The peak filling of \atom{87}{Rb}, $f_p$, as a function of number of \atom{87}{Rb} atoms with a lattice depth of $V_0=20E_R$ and harmonic confinements of $40\times40\times260$~Hz (filled triangles) and $35\times35\times227$ (filled circles) . Empty symbols are experimental data.  The crosses (``x" symbols) show the on-site density at the center of the trap at $T/J=15$ for the weaker (stronger) harmonic confinement. The data shown in the two insets as well as the profiles in the right hand side panel correspond to the stronger harmonic confinement. In the top-left inset we show the entropy matching procedure used to determine the final temperature of the \atom{87}{Rb} after they have been loaded adiabatically into the lattice. The blue circles show the entropy of $\sim$3000 \atom{87}{Rb} atoms in the lattice, while the solid, orange line shows the entropy of a weakly interacting bose gas.  The inset on the bottom-right shows the number of \atom{87}{Rb} atoms in doubly occupied sites at $T/J=1$ and $T/J=15$ using filled squares and filled circles, respectively. On the right panel we show $n(x,y)$ at points marked $I$ and $II$ at $T/J=0.2$ and $T/J=15$. }
\caption{(Color online) \emph{Peak filling of } species A vs. atom number. (Main panel) The peak filling $f_p$ as a function of number of \atom{87}{Rb} atoms, used as an example of bosonic species A, with a lattice depth of $V_0=20E_{\mathrm{A}}$ given for harmonic confinements of $\omega_r(\omega_z)=2\pi\,40(260)$~Hz (filled triangles) and $2\pi\,35(227)$~Hz (filled circles). Empty diamonds are experimental data~\cite{Moses_Covey_15}.  The dashed (solid) line shows the on-site density at the center of the trap at $T/J_{\mathrm{A}}=15$ for the weaker (stronger) harmonic confinement. (Lower inset) Entropy matching procedure for \atom{87}{Rb} following adiabatic lattice loading with strong harmonic confinement.  Solid green line is weakly interacting gas and blue triangles are $\sim$3000 lattice-confined Rb. (Upper inset) Double occupancy at $T/J_{\mathrm{A}}=1$ (filled upside-down triangles) and $T/J_{\mathrm{A}}=15$ (crosses) for the strong harmonic confinement.  (Right panels) Integrated density $n(x,y)$ at points marked I and II at $T/J_{\mathrm{A}}=1$ (upper) and $T/J_{\mathrm{A}}=15$ (lower) for strong harmonic confinement.
}
\label{fig:fig1}
\end{figure}

%\begin{figure}[t]
%\includegraphics[trim={1cm 2.5cm 4cm 0cm}, clip,width=0.5\textwidth]{Fig1.pdf}
%\label{fig:fig1}
%\end{figure}

To  characterize the density as a function of atom number, we first match the entropy of the gas in the dipole trap to the one in the lattice and  find the final temperature $T_{\rm fin}/J_{\mathrm{A}}$. The bottom right inset of Fig.~\ref{fig:fig1} shows an example of this procedure for 3000 atoms.  The solid  green line shows the entropy of a weakly interacting Bose gas, which describes the A atoms in the harmonic trap. Blue triangles  denote the entropy of the A atoms after loading into the lattice, given by
\begin{equation}
\label{eq:Entropy}
S_f(T)=\frac{E(T)-E(0)}{T} + \int_0^T \frac{E(T)-E(0)}{T^2} \;dT,
\end{equation} where $E(T)$ is the system energy at temperature $T$.  Both $E(T)$ and $E(0)$ are directly measured in our QMC simulations. An example of the entropy matching procedure for an initial temperature $T_i=0.4T_c$, $T_c$ the critical temperature for Bose-Einstein condensation, is shown with dashed arrows.

% Next we determine the peak filling, $f_p$,
%defined as {\bf
%\begin{equation}
%\label{eq:PeakFilling}
%f_p=N_{\rm Rb}/\left[ \int d\vec r \rho (\vec r)\right],(***FIX***)
%\end{equation}
%where $f_p\rho (\vec r)$ is a functional fit to the density distribution of the atoms in the trap, $n(\vec r)$, at a given %temperature $T/J_{\mathrm{A}}$ ($k_B=1$, obtained from QMC simulations$. (DO WE FIT ALONG Z OR INTEGRATE IT?}

Next we determine the peak filling, $f_p$.
 To compare with the experimental procedure of extracting the peak filling, we first integrate along $z$ and  mimic the experimental imaging resolution by applying a Gaussian filter with a width of 4 sites to the integrated density. We fit the resulting density to a  Thomas-Fermi (TF) distribution  $n(x,y)=\frac 43 f_p \sigma_z \left[1-(x/\sigma_x)^2-(y/\sigma_y)^2\right]^{3/2}$.  The extracted $f_p$ values are plotted in Fig.~\ref{fig:fig1}  as a function of atom number, $N_{\rm A}$, after the adiabatic loading procedure, with filled circles (triangles) for two different trapping conditions. The filling at the center of the trap is displayed with dashed (solid) lines for the weaker (stronger) harmonic confinement at $T/J_{\mathrm{A}}=15$. The empty diamonds are the experimental results~\cite{Moses_Covey_15}. The rightmost panels of Fig.~\ref{fig:fig1} give examples of the resulting distributions integrated along the $z$-axis, $n(x,y)$, for $N_{\rm A}\sim4000$ (labeled I) and $N_{\rm A}\sim8000$ (labeled II). For these two values of $N_{\rm A}$ we show the distribution at $T/J_{\mathrm{A}}=1$ (top), which is the low-temperature result, followed by the integrated density distribution at $T/J_{\mathrm{A}}=15$ (bottom), which is closer to the experimental temperature.

It is worth noting that while the density at the center of the trap is strongly dependent on temperature, $f_p$ does not display a strong dependence on $T/J_{\mathrm{A}}$ within the experimentally-relevant temperature range. However, this does not mean that the efficiency of the formation of molecules is unaffected by temperature. This is evident if one probes the number of doubly occupied sites with increasing temperature.  Sites with double (or higher) occupancies do not result in molecule formation.  The upper inset of Fig.~\ref{fig:fig1} shows the double occupancy at $T/J_{\mathrm{A}}=1$ (inverted triangles) and $T/J_{\mathrm{A}}=15$ (crosses). In the low-temperature regime ($T/J_{\mathrm{A}}=1$) the first Mott shell extends to $N_{\rm A} \approx 5000 $ particles, after which a superfluid region forms at the center of the trap, before transitioning to the second Mott shell at $N_{\rm A}\approx 8000$ (see the dashed blue line in Fig~\ref{fig:fig1}). However close to or above $T/J_{\mathrm{A}}=15$, $5-10\%$ of the A atoms are in doubly occupied sites and hence will not participate in molecule formation.

Next, we consider the fermionic species B of the mixture, taking \atom{40}{K} for comparison with recent experimental data~\cite{Moses_Covey_15}.  However, as before, our conclusions are valid for generic fermionic species that can be described by a tight-binding model.  The difference in the polarizability with respect to \atom{87}{Rb} means that the \atom{40}{K} atoms feel a lattice potential of depth $V_0=9E_{\mathrm{B}}$, where $E_{\mathrm{B}}$ is the B species recoil energy.  At low temperatures, $p$-wave interactions in the spin-polarized gas can be neglected, and the determination of the density reduces to a non-interacting problem.  As the harmonic trap is separable, this problem can be straightforwardly treated by direct diagonalization for the single-particle eigenenergies $E_{n_{\xi}}$ and corresponding wavefunctions $\vert \psi_{n_{\xi}}\rangle$. With these single-particle quantities we can evaluate the grand canonical partition function, from which we find the entropy per particle $S/N$, as well as the on-site density $n(\vec r)=n(x) n(y)  n(z)$,
\begin{equation}
\textstyle n(\xi)=\sum_{n_\xi} \frac{1}{1+\exp(\beta(E_{n_\xi}-\mu))} \vert\psi_{n_{\xi}}(\xi)\vert^2\, ,
\end{equation}
where $\xi=x,y,z$ and $\beta=1/T$ is the inverse temperature.

\begin{figure}[h]
\includegraphics[trim={0cm 4cm 8cm 2cm}, clip,width=0.5\textwidth]{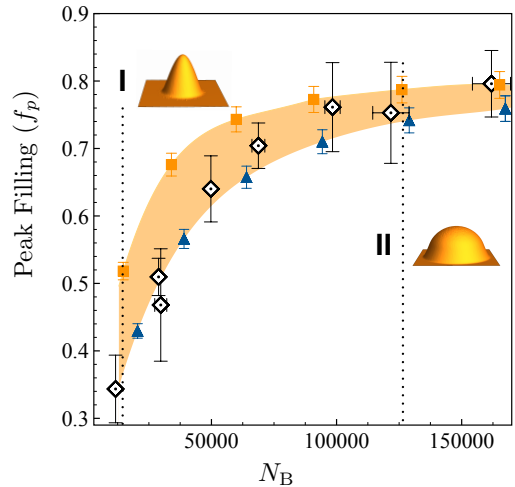}
\caption{(Color online) \emph{Peak filling of the fermionic species B vs.~atom number}. (Main panel) The peak filling of \atom{40}{K}, used as an example for the fermionic species B, at $T_{\rm fin}/J_{\mathrm{B}}=1$ (filled squares) and $T_{\rm fin}/J_{\mathrm{B}}=5$ (filled triangles) extracted from the integrated density distributions $n(x,y)$, including mimicked resolution and pixelation effects.  Results are for harmonic trap frequencies $(40\times40\times260)$~Hz and a $9E_{\mathrm{B}}$ deep lattice. Empty diamonds show experimental results~\cite{Moses_Covey_15}. Insets show the integrated density distributions at points I and II .  }
\label{fig:KPF}
\end{figure}

In Fig.~\ref{fig:KPF} we show the peak filling of species B as a function of atom number.   For fermions we use a Gaussian fit with $n(x,y)=\sqrt{2\pi \sigma_z} f_p\exp \left[-x^2/(2 \sigma_x^2)-y^2/(2 \sigma_y^2)\right]$ after  integration along $z$. Here, we also account for the experimental imaging resolution and pixelation by applying a Gaussian filter. The band, delimited by the blue triangles and orange squares, shows the range of peak fillings for $1<T_{\rm fin}/J_{\mathrm{B}}<5$, with $J_{\mathrm{B}}$ the tunneling of the B atoms. As the temperature of the B atoms in the lattice increases, so does the width of the cloud, resulting in a decrease in the peak filling. The empty diamonds show the experimental data~\cite{Moses_Covey_15}. The plateau indicates the atoms forming an incompressible band insulator. On the right panel we show the integrated density distribution at points marked I and II. At II, the B atoms form a band insulator in a large region of the lattice. 

{\it Lattice-confined two-species mixture }
We use a two-component mixture of soft-core (``A") and hard-core (``B") bosons to study the combined effect of interspecies interactions and finite temperature on the density of the A species.  Hard-core bosons act as a stand-in for fermionic B species, as path-integral QMC cannot simulate fermions due to the sign problem.  While there is no direct  mapping from hard-core bosons to fermions for local observables  in three dimensions, we compared the hard-core and fermionic profiles for the single-species case and found excellent agreement.  Based on this, we expect our results for the local density in the mixture to also be valid.

\begin{figure}[t]
\includegraphics[trim={1cm 3cm 5cm 4cm}, clip,width=0.5\textwidth]{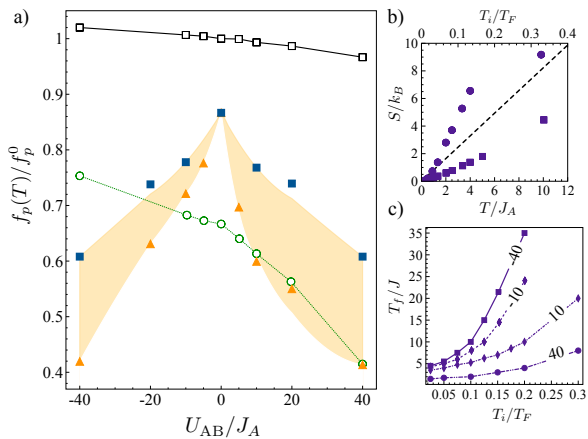}
\caption{(Color online) \emph{Dependence of species A peak filling on $U_{\mathrm{AB}}$}. (a) Peak filling normalized to zero-temperature value, $f_p(T)/f_p^0$, vs.~$U_{\rm AB}/J_{\mathrm{A}}$ at $T/J_{\mathrm{A}}$=0.2 (empty squares) and $T/J_{\mathrm{A}}$=10 (empty circles), ignoring the interaction dependence of the final temperature due to adiabatic loading.  Accounting for interaction effects during loading at initial temperatures $T_i/T_F=0.1$ (0.3) are given with filled squares (triangles). (b) The entropy of species A as a function of $T/J_{\rm A}$ is shown for $U_{\rm AB}/J_{\mathrm{A}}=40$ and -40 using circles and squares, respectively. The dashed line is the entropy of non-interacting fermions in a harmonic trap, which we use as an estimate for $S_i$. (c) Final temperature vs.~initial temperature for $U_{\rm AB}/J_{\mathrm{A}}=-40$, -10, 10, and 40 (solid, dash, dot-dash, and dot-dot-dash, respectively).  We use  a harmonic confinement of $\omega_r(\omega_z)=2\pi 23(150)$~Hz}
\label{fig:KRbPF}
\end{figure}

The two-component mixture is described by the Hamiltonian
\begin{align}
\label{eq:H}
\nonumber H&=-\sum_{\langle i\,j\rangle, \gamma }J_\gamma(a_{i,\gamma}^\dagger a_{j,\gamma}+\mathrm{h.c.})+\frac{U}{2}\sum_{i}n_{i,\rm A}(n_{i,\rm A}-1)\\
&+U_{\rm AB}\sum_i n_{i, \rm A} n_{i,\rm B}-\sum_{i, \gamma}\mu_{\gamma,i} n_{i,\gamma}.
\end{align}
Here $\gamma=$A, B and $U_{\rm AB}$ is the interspecies on-site interaction which can be tuned to be repulsive or attractive.  %In the following we consider a harmonic confinement of $(23\times 23\times 150)$~Hz. The A atoms feel a lattice depth of $V_0=20E_{\mathrm{A}}$, while the B species experience a shallower lattice potential of $V_0=9E_B$.
To study the effect of interspecies interactions we have performed simulations with $-40<U_{\rm AB}/J_{\mathrm{A}}<40$ with $10500$ A atoms and $25000$ B atoms. Previous studies have shown that at a given $T/J_{\mathrm{A}}$, the presence of attractive interactions enhances the conversion efficiency ~\cite{Freericks_efficiency}. This corresponds to an enhancement of $f_p$ measured at the same $T/J_{\mathrm{A}}$ for every $U_{\rm AB}<0$. In Fig.~\ref{fig:KRbPF} we show $f_p(T/J_{\mathrm{A}}=0.2)$ and $f_p(T/J_{\mathrm{A}}=10)$ as a function of $U_{\rm AB}/J_{\mathrm{A}}$ using filled squares and filled circles connected by a line, respectively. We normalize the values to $f_p^0$, the peak filling of A at $T/J_{\mathrm{A}}=0.2$ in the absence of species B. It is clear that attractive interactions lead to an enhancement of $f_p$ if $T/J_{\mathrm{A}}$ is kept constant.

However, this simple analysis does not describe current experiments, since starting at the same initial temperature, $T_i/T_F$, $T_F$ the Fermi temperature, leads to a final temperature $T_{\rm fin}/J_{\mathrm{A}}$ after adiabatic loading that depends on the sign and magnitude of $U_{\rm AB}$. In Fig.~\ref{fig:KRbPF} (c) we show $T_{\rm fin}/J_{\mathrm{A}}$ for $U_{\rm AB}/J_{\mathrm{A}}=-40$, -10, 10, and 40 using solid, dash, dot-dash, and dot-dot-dash lines, respectively. From Fig.~\ref{fig:KRbPF}(c) it is clear that attractive interactions tend to cause more severe heating during the adiabatic loading. For the range of initial entropies considered in this work, this additional heating inhibits and counteracts the benefit gained by making the two species attract. In Fig.~\ref{fig:KRbPF} (a) we show $f_p$ as a function of $U_{\rm AB}/J_{\mathrm{A}}$ for $T_i/T_F=$0.1 and 0.3 using filled squares and filled triangles, respectively.  Following adiabatic loading, both attractive and repulsive interactions destabilize the A species Mott insulator and reduce the peak filling.

{\it STIRAP adiabaticity and higher-band transfer.}
Pairs of A and B atoms in an optical lattice site are transferred to weakly bound Feshbach molecules following a magneto~\cite{Ni_Ospelkaus_08}- or photo~\cite{PhysRevA.89.020702}-association step.  These Feshbach molecules are converted into ground state molecules via a two-photon STIRAP sequence involving an intermediate electronically excited molecular state.  As the STIRAP process must remove $\sim100$ THz of molecular energy, the difference of the wavevectors of the two photons involved in the STIRAP sequence, denoted by $\vec{k}_u$ and $\vec{k}_d$, can have significant variation on the lattice scale of a few microns. The resulting momentum transfer can excite the resulting ground state molecules to higher lattice bands.  Here, we investigate the adiabaticity of the STIRAP procedure, as well as the rate of transfer of molecules to higher bands.

We consider a STIRAP process which couples the Feshbach molecule (FBM) ($\vert f \rangle$) and ground molecular (GSM) $\vert g \rangle$ states through an excited level ($\vert e \rangle$)~\cite{Ospelkaus_Peer_08}. We assume that these are the only states involved and neglect the population of other molecular levels or atomic scattering states during the STIRAP.  To match current experiments, we study the case of vanishing single- and two-photon detunings.  Using adiabatic perturbation theory~\cite{Polkovnikov_APT, Rigolin_APT,Supp} we find that the momentum transfer following the process $|f\rangle\to|g\rangle$ has the same form as the case of large single-photon detuning, $\propto e^{i (\vec k_u -\vec k_d)\cdot\vec{r}}$. Typical STIRAP linewidths are $\sim200$kHz, significantly greater than the $\sim20$kHz spacing between bands, and so the band structure is fully unresolved. Thus, the relative population of molecules in the different bands following STIRAP are determined by ratios of Rabi frequencies in the basis of Wannier states with band index $n$. We use $w_n(\vec r)$ and $\bar w_n(\vec r)$ to denote the Wannier states for FBM and GSM, respectively. For FBM in the lowest band, the relative population of GSM in the first excited band along the direction $\xi$ can thus be estimated by
\begin{align}
\left|\frac{\langle \bar w_1|e^{ik_{\xi} \xi}|w_0\rangle}{\langle \bar w_0|e^{ik_{\xi} \xi}|w_0\rangle}\right|^2&\approx \frac{2 (k_{\xi}a)^2\tilde{\alpha}}{\pi^2(1+\tilde{\alpha})^2\sqrt{V/E_R}}\, ,
\end{align}
where we have used a harmonic approximation for the Wannier functions, with $k_{\xi}\equiv (\vec k_u -\vec k_d)\cdot\vec{\xi}$ the momentum transfer along direction $\xi$, $a$ the lattice spacing, $\tilde{\alpha}$ the polarizability ratio of the GSM to the FBM, $V$ the lattice depth for the FBM, and $E_R$ the molecular recoil energy. For the parameters of the JILA experiment~\cite{Moses_Covey_15}, $a=532$~nm, $k_u=2\pi/(968$~nm) and $k_d=2\pi /(689$~nm) co-propagating at a 45$^{\circ}$ angle with respect to the $x$ and $y$ lattice axes, and $\tilde{\alpha}\approx 0.9$, we find the total population in the first excited bands to be $\sim1\%$. In general, our results indicate that for experiments in the Lamb-Dicke regime $k_{\xi} a/(V/E_R)^{1/4}\ll 1$, STIRAP does not induce appreciable population transfer to excited bands.

In conclusion, we studied the combined effects of interspecies interactions, temperature, and adiabatic loading on the successful preparation of a low entropy gas of polar molecules. We have shown that interspecies interactions have a significant effect on the final temperature of the lattice gas following adiabatic loading, which in turn {can} lead to the depletion of the peak filling of the bosonic species and a lower efficiency of molecule formation.  Based on our results, molecule formation efficiency is greatest when the lattice loading is performed with a non-interacting mixture.  For lower initial temperatures where the final temperature after loading is $ T_{\rm fin}/J_{\mathrm{A}}\lesssim 5$, attractive interactions may lead to a higher rate of molecule formation; this temperature is around $T/T_F\lesssim 0.05$ for the parameters considered in this work.  Moreover, we considered the role of the STIRAP procedure used to convert the resulting Feshbach molecules to ground state molecules in the conversion efficiency. We find that the probability of promoting molecules to higher bands during STIRAP is not a significant concern provided the Lamb-Dicke criterion is satisfied. 

\emph{Acknowledgments} We would like thank B. Capogrosso-Sansone for useful discussions. This
work was supported by the NSF (PIF-1211914 and PFC-
1125844), AFOSR, AFOSR-MURI, NIST and ARO individual
investigator awards.  MLW thanks the NRC for support.  Part of the computing for this project was performed at the OU Supercomputing Center for Education $\&$ Research (OSCER) at the University of Oklahoma (OU). This work also utilized the Janus supercomputer, which is supported by the National Science Foundation (award number CNS-0821794) and the University of Colorado Boulder. The Janus supercomputer is a joint effort of the University of Colorado Boulder, the University of Colorado Denver and the National Center for Atmospheric Research.

%\bibliography{quasi}

\end{document}